\documentstyle[epsfig]{mn}

\newif\ifAMStwofonts
\AMStwofontstrue

\newcommand{\be}{\begin{equation}}
\newcommand{\ee}{\end{equation}}

\ifoldfss
  \ifCUPmtlplainloaded \else
    \NewTextAlphabet{textbfit} {cmbxti10} {}
    \NewTextAlphabet{textbfss} {cmssbx10} {}
    \NewMathAlphabet{mathbfit} {cmbxti10} {} 
    \NewMathAlphabet{mathbfss} {cmssbx10} {} 
  \fi
  \ifAMStwofonts
    \ifCUPmtlplainloaded \else
      \NewSymbolFont{upmath} {eurm10}
      \NewSymbolFont{AMSa} {msam10}
      \NewMathSymbol{\upi}     {0}{upmath}{19}
      \NewMathSymbol{\umu}     {0}{upmath}{16}
      \NewMathSymbol{\upartial}{0}{upmath}{40}
      \NewMathSymbol{\leqslant}{3}{AMSa}{36}
      \NewMathSymbol{\geqslant}{3}{AMSa}{3E}

    \fi
  \fi
\fi 

\ifnfssone
  \newmathalphabet{\mathit}
  \addtoversion{normal}{\mathit}{cmr}{m}{it}
  \addtoversion{bold}{\mathit}{cmr}{bx}{it}
  \newmathalphabet{\mathbfit} 
  \addtoversion{normal}{\mathbfit}{cmr}{bx}{it}
  \addtoversion{bold}{\mathbfit}{cmr}{bx}{it}
  \newmathalphabet{\mathbfss} 
  \addtoversion{normal}{\mathbfss}{cmss}{bx}{n}
  \addtoversion{bold}{\mathbfss}{cmss}{bx}{n}
  \ifAMStwofonts
    \ifCUPmtlplainloaded \else
      \UseAMStwoboldmath
      \makeatletter
      \new@mathgroup\upmath@group
      \define@mathgroup\mv@normal\upmath@group{eur}{m}{n}
      \define@mathgroup\mv@bold\upmath@group{eur}{b}{n}
      \edef\UPM{\hexnumber\upmath@group}
      \new@mathgroup\amsa@group
      \define@mathgroup\mv@normal\amsa@group{msa}{m}{n}
      \define@mathgroup\mv@bold\amsa@group{msa}{m}{n}
      \edef\AMSa{\hexnumber\amsa@group}
      \makeatother
      \mathchardef\upi="0\UPM19
      \mathchardef\umu="0\UPM16
      \mathchardef\upartial="0\UPM40
      \mathchardef\leqslant="3\AMSa36
      \mathchardef\geqslant="3\AMSa3E
    \fi
  \fi
\fi 

\ifnfsstwo
  \DeclareMathAlphabet{\mathbfit}{OT1}{cmr}{bx}{it}
  \SetMathAlphabet\mathbfit{bold}{OT1}{cmr}{bx}{it}
  \DeclareMathAlphabet{\mathbfss}{OT1}{cmss}{bx}{n}
  \SetMathAlphabet\mathbfss{bold}{OT1}{cmss}{bx}{n}
  \ifAMStwofonts
    \ifCUPmtlplainloaded \else
      \DeclareSymbolFont{UPM}{U}{eur}{m}{n}
      \SetSymbolFont{UPM}{bold}{U}{eur}{b}{n}
      \DeclareSymbolFont{AMSa}{U}{msa}{m}{n}
      \DeclareMathSymbol{\upi}{0}{UPM}{"19}
      \DeclareMathSymbol{\umu}{0}{UPM}{"16}
      \DeclareMathSymbol{\upartial}{0}{UPM}{"40}
      \DeclareMathSymbol{\leqslant}{3}{AMSa}{"36}
      \DeclareMathSymbol{\geqslant}{3}{AMSa}{"3E}
    \fi
  \fi
\fi 

\ifCUPmtlplainloaded \else
  \ifAMStwofonts \else 
    \def\upi{\pi}
    \def\umu{\mu}
    \def\upartial{\partial}
  \fi
\fi

\begin{document}

\title[Topological Statistics]{\bf Topological Statistics and the LMT Galaxy 
	Redshift Project}
 
\author[D.  Mitsouras et al.]{Dimitris Mitsouras$^1$, 
	Robert Brandenberger$^{2,3}$ and Paul Hickson$^3$\\ 
	$^1$Computer Science Department, Brown University,
	Providence, R.I.  02912, USA\\ 
	$^2$Physics Department, Brown University,
	Providence, R.I.  02912, USA\\ 
	$^3$Department of Physics and Astronomy,
	University of British Columbia, Vancouver, B.C.  V6T 1Z1, CANADA} 
\maketitle

\begin{abstract}  

\noindent The results of numerical simulations are presented which demonstrate
that liquid mirror telescope galaxy redshift surveys such as the current
UBC-NASA Multi-Narrowband Survey and the future LZT Survey have the potential of
discriminating between the predictions of different theories of structure
formation.  Most of the currently studied theories of structure formation
predict a scale-invariant spectrum of primordial perturbations.  Therefore, to
distinguish between the predictions of the various models, we make use of
statistics which are sensitive to non-Gaussian phases, such as the counts in
cell statistics, N-galaxy probability functions and Minkowski functionals.  It
is shown that already the current UBC-NASA survey can clearly differentiate
between the predictions of some topological defect theories and those of
inflationary Universe models with Gaussian phases.

\end{abstract}
\begin{keywords}
cosmology: theory -- large-scale structure, redshift surveys
\end{keywords}

\noindent BROWN-HET-1116\hfill April 1998.
 
\section{Introduction} 

Liquid mirror telescope (LMT) redshift surveys provide an efficient way to study
the distribution of galaxies in the Universe.  Recent technological developments
at Laval University and the University of British Columbia (Borra et al.  1992,
Hickson et al.  1994) have made liquid mirrors a viable low-cost alternative to
glass mirrors for zenith-pointing telescopes.  These surveys employ a set of
medium-band filters to measure the spectral energy destributions (SEDs) of every
detectable object in a long strip of sky a few tenths of a degree wide.  From
the SEDs, galaxy types and redshifts are estimated by fitting spectral templates
to the data.  The resulting redshifts have typical redshift accuracy of order
0.01 - 0.03 depending on the wavelength coverage and signal-to-noise ratio
(Hickson, Gibson and Callaghan 1994).  The technique is similar to that of
photometric redshifts derived from broad-band filters (eg.  Koo 1985, Lanzetta
et al 1996) but provides about an order of magnitude improvement in redshift
accuracy for the same signal-to-noise ratio.  A first such survey, the UBC-NASA
Multi-Narrowband survey (Hickson \& Mulrooney, 1998), is nearing its completion,
and a new larger-scale project, the LZT survey, which will probe the Universe to
a much greater depth, is in construction.

The 6-m Large Zenith Telescope (LZT) (Hickson et al.  1998) is expected to
provide redshifts for galaxies as faint as 23rd magnitude.  This is several
magnitudes fainter than can be reached by current and planned wide-angle
spectroscopic surveys such as the 2DF (Maddox 1998) and Sloan Digital Sky
Survey.  However, the photometrically derived redshifts of the UBC-NASA survey
and future LZT are less accurate than spectroscopic redshifts.

The goal of this study is to determine if, in spite of the substantial redshift
errors and of the narrow-strip geometry of the survey region, liquid mirror
galaxy redshift surveys have the potential to reveal important information about
the origin of structure in the Universe.  Specifically, we are interested in
exploring whether the surveys can distinguish between the predictions of models
of structure formation which are characterized by the same primordial power
spectrum of mass fluctuations, but which have different phase relations, e.g.
on one hand inflationary Universe models with Gaussian random phases, and on the
other hand topological defect theories characterized by rich nontrivial phase
correlations.

A second goal of this study is to explore which aspects of the technology future
improvement should focus on if the goal of the survey is to gain more
information about the origin of structure in the Universe.  Specifically, what
does one gain by increasing the depth of the survey compared to what one would
gain by increasing the accuracy of the redshifts by adding more filters.

We study these issues by performing numerical simulations of various theories of
structure formation, slicing from the simulation box regions with the geometry
of the LMT surveys, and analyzing the results using statistical measures which
are sensitive to a non-Gaussian distribution of phases.  Specifically, we shall
evaluate counts-in-cell (CIC) statistics, N-galaxy probability functions (NPF)
and Minkowski functionals (Mecke et al.  1994).

We compare the results of inflation-based cold dark matter (CDM) theories and
topological defect models.  Since the dynamics of actual defect models is very
complicated and not yet completely understood, we will investigate toy model
realizations constructed to maximize the degree of non-Gaussianness.  We study
cosmic string and global texture models (a global monopole theory would be
rather similar to the texture model).  For strings, we study two extreme
versions of the theory.  The first is a ``filament" model which is realized if
the strings have a lot of small-scale structure and therefore are characterized
by a small transverse velocity, leading to a filamentary pattern of accretion.
The second is a ``wake" model which is realistic if the strings have little
small-scale structure, large transverse velocity, and no local gravitational
potential, and thus lead to planar velocity perturbations in their wake.

Our simulations demonstrate that already with the current UBC-NASA survey useful
lessons for cosmology can be deduced.  With a limiting magnitude (AB magnitude
scale) of 20.4 and galaxy redshift errors of $\Delta z < 0.03$ (based on the use
of 33 filters), the complete survey will allow us to differentiate between the
predictions of idealized defect theories and Gaussian models.  With an
improvement of the redshift errors and survey depth which will be possible with
the LZT survey, the distinguishing power of the survey will be greatly
increased.

\section{Models}

\noindent{\it Standard Cold Dark Matter Model}

There are two main classes of theories of structure formation which have been
investigated over the past decade.  The first are inflation-based models with a
scale-invariant spectrum of primordial adiabatic perturbations with random
phases, each having a Gaussian distribution.  These models are viable only if
the dark matter is cold (resulting in the so-called {\it Cold Dark Matter} (CDM)
models).  In this paper, we shall study the ``standard" CDM model with an
exactly scale-invariant spectrum of perturbations, $\Omega = 1$, and vanishing
cosmological constant.  This specific model is no longer viable if we compare
the normalization of the power spectrum of density perturbations with that of CMB
fluctuations.  Viable CDM models either require an admixture of hot dark matter,
a value of $\Omega$ smaller than one, a remnant cosmological constant
contributing significantly to $\Omega$, or a tilt of the primordial spectrum of
perturbations.  However, in this paper we are mostly concerned with the power of
the LMT survey s to pick up signatures of primordial non-random phases, and the
model we study is taken as the canonical model with Gaussian random phases.

Thus, we consider a CDM model with $\Omega = 1$, $h = 0.5$, with vanishing
cosmological constant, and having an exactly scale-invariant primordial spectrum
of cosmological perturbations (generated as quantum fluctuations in an early
period of inflation (Chibisov \& Mukhanov 1980, Chibisov \& Mukhanov 1982,
Lukash 1980, Mukhanov \& Chibisov 1981, Guth \& Pi 1982, Hawking 1982,
Starobinskii 1982, Bardeen et al.  1983).  The processing of the primordial
spectrum of perturbations at late times on scales smaller than the Hubble radius
leads to a nontrivial transfer function.  We take the transfer function given in
Bardeen et al.  (1986).

A cosmological model with these parameters was evolved numerically in a box of
comoving length of $700$Mpc with $128^3$ particles.  The normalization of the
primordial power spectrum was chosen such that $\sigma_8 = 0.95$ (note that a
COBE normalized model would require $\sigma_8 = 1.22$.  The numerical work was
done with the Bertschinger \& Gelb (1991) $P^3M$ N-body code and was run on the
CRAY-YMP at the Brown Center for Scientific Computation.

\noindent{\it String Wake Model}

The second class of structure formation models which has received a lot of
attention over the past decade is based on topological defects giving rise to
the seeds for structures (see Vilenkin \& Shellard (1994), Hindmarsh \& Kibble
(1995) or Brandenberger (1994) for recent reviews).  Models based on strings,
global monopoles, and textures are of special interest (models with local
monopoles or domain walls are ruled out since the defects in these models would
overclose the Universe).

Topological defects are a generic prediction of a wide class of particle physics
models.  If the set of vacuum states of the theory after symmetry breaking has a
nontrivial topology, then defects will inevitably be produced during the
symmetry breaking phase transition in the early Universe (Kibble 1976).  The
defects form non-adiabatic density perturbations and act as the seeds of
structure formation.  If the scale of symmetry breaking is about $10^{16}$GeV,
then the resulting spectrum of perturbations has the correct amplitude to
explain the large-scale fluctuations in the cosmic microwave background observed
by COBE (Smoot et al.  1992).

Cosmic strings are linear topological defects which arise if the vacuum manifold
of the particle physics theory is not simply connected.  At the time of the
phase transition, a network of strings is formed.  The strings are either
infinite in length or closed loops.  The {\it long} string network consists of
all the strings whose curvature radius is greater than the Hubble radius.
Causality considerations show that the correlation length $\xi(t)$ of the long
string network (the mean separation of the long strings) must be smaller or
equal to the Hubble radius at all times after the phase transition.  Analytical
arguments (Vilenkin 1985) and numerical studies (Albrecht \& Turok 1989, Bennett
\& Bouchet 1988, Allen \& Shellard 1990) demonstrate that the string network
takes on a {\it scaling solution} in which $\xi(t) \sim t$ for all $t$.  Hence,
it follows that the cosmic string model predicts a scale-invariant spectrum of
primordial perturbations, much like inflationary Universe models.  In contrast
to the latter, however, the phases of the perturbations in the string model are
non-random.

The precise nature of the non-random phases (more specifically the topology of
the dominant structures produced in the cosmic string model) depend on some
details of the string dynamics which are, unfortunately, not yet resolved.  If
the strings are straight on the scale of the Hubble radius, then the string
tension will equal the mass per unit length $\mu$.  As a consequence, the
strings will move relativistically, and they exert no local gravitational force.
In contrast, if the long strings have a lot of small-scale structure, then the
tension will be smaller than $\mu$, the transverse motion of the strings will be
slower and there will be a local gravitational force exerted by the strings.
The first scenario gives rise to the {\it cosmic string wake model}, the second
to the {\it cosmic string filament model}.

Note that since the strings are non-adiabatic seed perturbations, then, in a
model in which the dark matter is hot, free streaming does not erase the
primordial perturbations on small scales.  Hence, the cosmic string model of
structure formation is viable even if the dark matter is hot (Vilenkin \& Shafi
1983, Brandenberger et al.  1987a, 1987b).

A straight cosmic string without small-scale structure generates a conical space
perpendicular to the string with deficit angle (Vilenkin 1981)
\begin{equation}
\alpha \, = \, 8 \pi G \mu  \, 
\end{equation}
(We use units in which the speed of light $c = 1$). If the string is moving 
through a plasma with transverse velocity $v$, this induces a velocity 
perturbation (Silk \& Vilenkin 1984) of magnitude
\begin{equation}
\delta v \, = \, 4 \pi G \mu \gamma(v) \, ,
\end{equation}
where $\gamma(v)$ is the relativistic $\gamma$ factor associated with the 
velocity $v$. The velocity perturbations, in turn, leads to a non-adiabatic 
density perturbation of roughly planar geometry - a wake. The dimensions of 
the plane are determined by the string velocity $v$ and by the correlation 
length $\xi(t)$. A string at time $t_i$ produces a wake of comoving length 
$v \xi(t_i) z(t_i)$ and width $\xi(t_i) z(t_i)$. The wake thickness can be 
calculated (Perivolaropoulos et al. 1990, Brandenberger 1991) using the 
Zel'dovich approximation (Zel'dovich 1970) as the comoving scale of the 
region which has separated from the Hubble flow and is no longer expanding. 
Because of free streaming, the result depends on the nature of the dark 
matter. For hot dark matter, the thickness at the present time $t_0$ of a 
wake created at time $t_i > t_{eq}$ is (Perivolaropoulos et al. 1990, 
Brandenberger 1991)
\begin{equation}
h(t_i) \, = \, {{24 \pi} \over 5} G \mu v \gamma(v) z(t_i)^{1/2} t_0 \, .
\end{equation}
Thus, the wake dimensions are
\begin{equation}
v \xi(t_i) z(t_i) \, \times \, \xi(t_i) z(t_i) \, \times h(t_i) \, .
\end{equation}
The mass of such a structure is
\begin{equation}
m(t_i) \, = \, {{24 \pi} \over 5} G \mu v^2 \gamma(v) \nu_1^{-2} 
\rho_0 t_i^{1/3} t_0^{8/3} \, ,
\end{equation}
where we have taken the string correlation length to be 
$\xi(t) = {1 \over \nu_1} t$.

Hot dark matter wakes created before the time $t_{eq}$ of equal matter and
radiation are diluted by the radiation pressure and will not be included in our
simulations.  Note that in the cosmic string wake model, the structures which
dominate the large-scale distribution of galaxies are predicted to be planar
wakes.  The most numerous and thickest are those created at $t_{eq}$.  Their
size is comparable to that of the Great Wall (de Lapparent et al.  1986).

The procedure for setting up a toy wake model simulation is the following.  The
time interval between $t_{eq}$ and $t_0$ is divided into Hubble times.  For the
cosmological parameters used in this paper, there are 16 Hubble time steps.  At
each time step, labeled by a time $t_i$, a number $c_w^{-3}$ of wakes are laid
down per volume $t_i^3$.  Here, $c_w t_i$ gives the mean separation of wakes at
this time.  Wakes are laid down by selecting centers and orientations of the
wakes at random within each volume $t_i^3$.  For given values of $G \mu$, of the
mean transverse velocity of the string $v$ and of the number $\nu_1$, the value
of $c_w$ can be determined by demanding that the total mass density adds up to
critical density.

In principle, the number $\nu_1$ could be determined by numerical simulations.
In practice, however, the resolution of the cosmic string evolution codes may
not be good enough to give a reliable answer.  Thus, we will consider a range of
values.  To be specific, we choose the following values:  $G \mu = 10^{-6}$, $v
\gamma(v) = 1$ and $\nu_1$.

Given the geometrical distribution of wakes, galaxies are laid down at random
within the volume of the wakes, with a uniform density.

\noindent{\it String Filament Model}

Long cosmic strings with a substantial amount of small-scale structure (e.g. 
small-scale wiggles) will have a coarse-grained equation of state which 
differs from that of a relativistic string, i.e. $|p| < \mu$, where $|p|$ 
denotes the string tension. In this case, the strings will have a small 
transverse velocity, but they will exert a local gravitational attraction, 
and hence they will seed filamentary density perturbations. We shall take the 
extreme situation in which the transverse string velocity is negligible (the 
intermediate case was analyzed in Zanchin et al. (1996)). In this case, the 
comoving radius $q_{nl}(t_i)$ of a filament seeded at time $t_i$ can again 
be calculated using the Zel'dovich approximation, yielding (Aguirre \& 
Brandenberger 1995) 
\begin{equation}
q_{nl}(t_i) \, = \, \bigl({{12 G \lambda} \over 5} {\rm ln}({{t_0} \over {t_i}})\bigr)^{1/2} t_0 \, ,
\end{equation}
where $\lambda = \mu - |p|$. Hence, the mass in an individual filament which 
was seeded at time $t_i$ is
\begin{equation}
m(t_i) \, = \, {{12 \pi G \lambda} \over 5} {1 \over \nu_1} \rho_0 t_0^{8/3} 
t_i^{1/3} {\rm ln}\bigl({{t_0} \over {t_i}}\bigr) \, ,
\end{equation}
where $\nu_1$ is defined as in the previous subsection.

The procedure for setting up the string filament simulation is analogous to how
the wake toy model is constructed.  At each Hubble time step between $t_{eq}$
and $t_0$, a fixed number $c_f^{-3}$ of filaments for volume $t^3$ is laid down,
choosing filament centers and directions at random.  For given values of $G
\lambda$ and $\nu_1$, the number of filaments is determined by demanding that
the total mass adds up to critical density.  Galaxies are again put down at
random in the volume of the filaments, with a uniform density.  To be specific,
we shall consider the values $G \lambda = 10^{-6}$ and $\nu_1 = 1$.

Note that the string filament and string wake models represent opposite extremes
of what we expect a realistic string model to look like.  The statistical study
of the wake and filament models will hence yield as a byproduct a quantitative
measure of the large uncertainty in the predictions of the string model, in
particular if we let $\nu_1$ vary over a significant range of values.

\noindent{\it Texture Model}

Textures (Davis 1987) are topological defects which arise if the vacuum manifold
of the particle physics model has nontrivial third homotopy group.  In four
space-time dimensions, textures are unstable (Turok 1989).  Texture
configurations on Hubble scale unwind within approximately one Hubble expansion
time.  Nevertheless, in theories with a global symmetry, textures are able to
generate density perturbations.  In field theories with a local symmetry,
texture configurations in the scalar matter fields can be compensated by gauge
fields on a microscopic time scale, and hence in such models textures are not
important for structure formation.

The simplest texture configurations (Turok 1989) are spherically symmetric.
Hence, we will construct a texture toy model by superimposing individual
spherically symmetric texture configurations.  Obviously, this is a gross
simplification of what a texture model will really look like.

At each time $t_i$, there is a finite probability (Prokopec 1991, Leese and 
Prokopec 1991) $p$ that in a volume $t_i^3$ there will be a texture 
configuration. The initial texture configuration can be modeled (Gooding 
et al. 1991) by a spherically symmetric velocity perturbation which then 
develops into a spherically symmetric density perturbations. The comoving 
radius $q_{nl}(t_i)$ of this perturbation can again be determined (Aguirre 
1995) by the Zel'dovich approximation, yielding the result
\begin{equation}
q_{nl}(t_i) \, = {6 \over 5} \epsilon t_i^{- 1/3} t_0^{4/3} \, ,
\end{equation}
where (up to factors of order unity) $\epsilon = G \eta^2$, $\eta$ being the 
scale of symmetry breaking. We will take the value $\epsilon = 3 \times 
10^{-5}$, an appropriate value for a COBE-normalized texture model (Pen 
et al. 1997). The mass in a texture can be computed by multiplying the volume 
of the texture by the background density.

The texture toy model was obtained by laying down $p$ texture centers per volume
$t_i^3$ at each Hubble expansion time $t_i$ with a random choice of the centers.
Galaxies were distributed within each texture with a Gaussian density profile
(as a function of radius) and random angular distribution.  To agree with the
total mass obtained in the other toy models, the number of galaxies per texture
was chosen to (for fixed values of $\epsilon$ and $p$) yield critical density
for the entire volume.

\noindent{\it General Comments}

Note that all four models discussed above predict a roughly scale-invariant
primordial spectrum of perturbations.  In the case of inflation, the reason for
the scale invariance can be explained by the basic geometry of an inflationary
cosmology (see e.g.  Press (1980) or Brandenberger (1985)), in defect models the
approximate scale-invariance is a direct consequence of the scaling dynamics of
the defect system (see e.g.  Brandenberger (1994)).  However, the morphology in
these four models is very different, and with the statistics discussed below we
are able to differentiate between their predictions for the large-scale
structure of the Universe.

\begin{figure}
\epsfxsize=3.3in \epsfbox{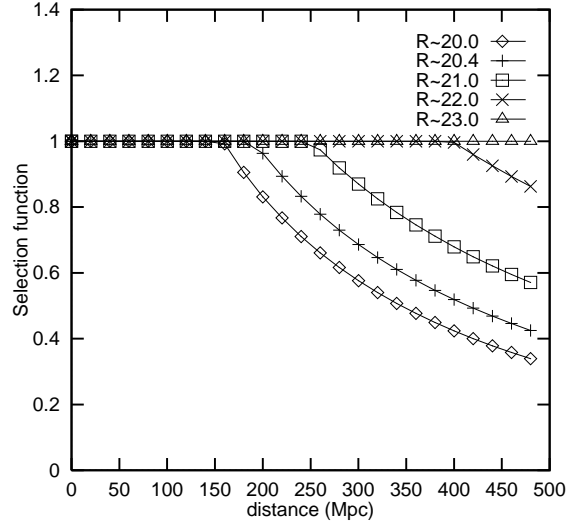}
\caption{The LMT selection function for various values of the limiting 
magnitude.}
\end{figure}

\section{Statistics}

\noindent{\it Counts in Cell Statistic}

The first statistic evaluated for our simulations was a simple {\it counts in
cell} (CIC) statistic (see e.g.  Saslaw (1989)).  Each LMT slice was projected
into a two-dimensional plane and then divided into equal volume cells
(three-dimensional volume).  An equal number of cells in radial and angular
direction was chosen.  Note that the shape of the cells depends strongly on the
redshift, i.e.  the cells are not isometric.  However, since we are comparing
the predictions of different models (and eventually with observations) all
treated the same way, the unusual cell shape distribution should not be a
problem.

The CIC statistic is a histogram of number of cells with $n$ galaxies as a
function of the number $n$.  The CIC statistic can be evaluated for any cell
size (smoothing length).  Since the difference in the predictions of the models
is expected to be most important on scales slightly less than the size of the
dominant structures expected in our models, i.e.  of comoving length scale
$t_{eq} z(t_{eq})$, a smoothing length comparable to this scale must be chosen.
We took 34 cells per side which corresponds to a cell length of about
$15h^{-1}$Mpc.

\noindent{\it N-Galaxy Probability Function}

The second measure we used is the N-galaxy probability function (NPF), a
generalization of the void probability function (Hamilton et al.  1985).  The
NPF (Aguirre 1995) of a galaxy distribution is the probability that there are
exactly $N$ galaxies in a randomly placed volume $V$, considered as a function
of $V$.  Instead of plotting the NPF's, we have found it more informative to
plot the cumulative NPF, which for fixed value of $N$ is the probability that
there are less than or equal to $N$ galaxies within a randomly placed volume
$V$.

\noindent{\it Minkowski Functionals}

The Minkowski functionals (Mecke and Wagner 1991) supply a complete 
characterization of the global morphology of an a galaxy distribution. Given 
an isodensity surface $\partial K$ in three spatial dimensions, the morphology 
is determined by four quantities. The first functional $M_0$ is the volume $V$ 
enclosed by the surface
\begin{equation}
M_0 \, = \, V \, ,
\end{equation}
the second is related to the surface area $A$ 
\begin{equation}
M_1 \, =\,  A/8 \, ,
\end{equation} 
the third is proportional to the integrated mean curvature $H$
\begin{equation}
M_2 \, = \, {1 \over {2 \pi^2}} H
\end{equation}
with
\begin{equation}
H \, = \, {1 \over 2} \int_{\partial K} dA \bigl( {1 \over {R_1}} + {1 \over
 {R_2}} \bigr) \,
\end{equation}
where $R_1$ and $R_2$ are the principal curvature radii of the surface. 
Finally, the fourth Minkowski functional is related to the Euler 
characteristic $\chi$ and hence to the genus
\begin{equation}
M_3 \, = \, {{3 \chi} \over {4 \pi}} \, ,
\end{equation}
where
\begin{equation}
\chi \, = \, \int_{\partial K} dA {1 \over {R_1 R_2}} \, .
\end{equation}

There are many ways of displaying the Minkowski functionals as a function for
the density contrast.  We use the method of Mecke et al.  (1994) and Kerscher et
al.  (1997) and display the functionals as a function of a length scale $r$.
Given a value of $r$, we consider a volume made up of the union of balls of
radii $r$ whose centers are the galaxies in our sample.  Note that $r$ play the
role of a neighborhood relation length and is similar to a smoothing length of
the density field.  The larger the value of $r$, the smaller is the value of the
density contrast whose isodensity surface the Minkowski functionals are
describing.  We made use of modified versions of codes to evaluate the Minkowski
functionals made available by Kerscher et al.  (1997) and Schmalzing \& Buchert
(1997), which include boundary corrections.

\begin{figure}
\epsfxsize=3.3in \epsfbox{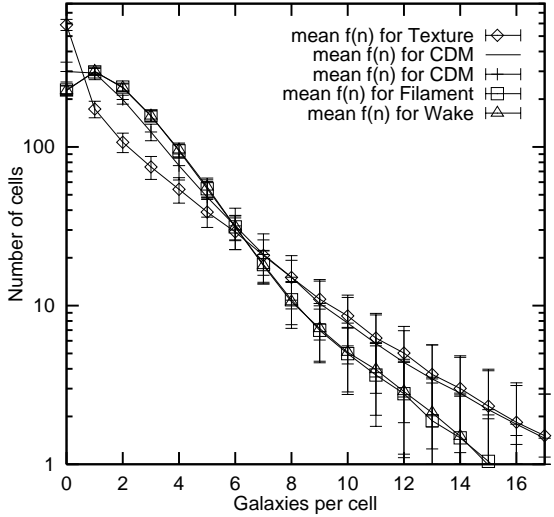}
\caption{The CIC statistic applied to the LMT geometry for simulations of the 
standard CDM model, the texture toy model, and the string wake and filament 
toy models. Redshift error $\Delta z = 0.03$ and limiting magnitude $R = 20.4$ 
correspond to the values of the current UBC-NASA survey.}
\end{figure}

\begin{figure}
\epsfxsize=3.3in \epsfbox{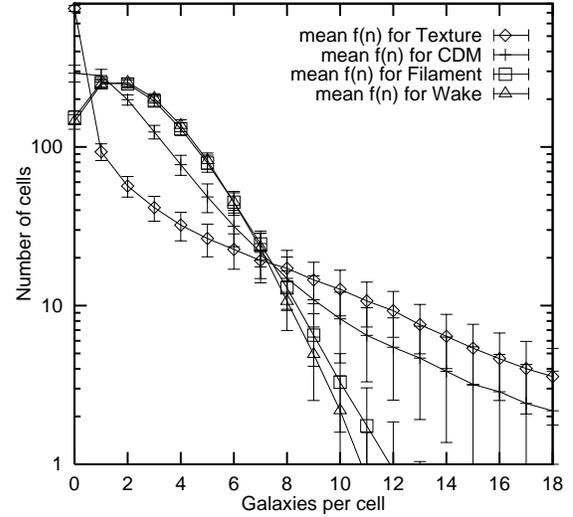}
\caption{Same as Figure 2, but for a survey with smaller redshift errors of 
$\Delta z = 0.01$.}
\end{figure}

\section{Simulations and Results}

Our analysis is done in the following way.  A simulation of each of the models
discussed in Section 2 is performed in a box of side length $700h^{-1}$Mpc.
Next, a slice with the geometry of the LMT survey ($90^o$ in length and $0.26^o$
in width) is cut out of this box.  The slice has a depth of $500h^{-1}$Mpc.
Thus, in order to compare observational results to the simulations, a volume
limited subset of the observational data must be used.

To compare the simulations with a survey with a limiting apparent magnitude 
$m_{limit}$, particles from the simulations are picked at random using a 
selection function $s(d)$, where $d$ is the distance from the origin. The 
selection function was determined by assuming a Press-Schechter luminosity 
function
\begin{equation}
f(L) \, = c \bigl({{L} \over {L^*}}\bigr)^{\alpha} e^{- L/L^*} \,
\end{equation}
where $L$ stands for the luminosity and $c$ is a constant which drops out of 
the analysis. The value of $L^*$ was taken (Zucca et al. 1997) to correspond 
to the absolute magnitude $M^* = - 19.6 + 5 {\rm log} h$, and a value of 
$\alpha = - 1.22$ appropriate for field galaxies was chosen. A sharp lower 
cutoff on this luminosity function at a value corresponding to $M = -16$ was 
used. The resulting selection function is shown in Figure 1 for various values 
of the limiting apparent magnitude.

To mimic the redshift errors expected in the LMT survey, each galaxy was moved
in redshift direction by a redshift error chosen from a Gaussian distribution
with standard deviation $\Delta z$.  For a fixed box, the processes of picking
an LMT slice, selecting galaxies and moving their redshifts randomly as
described above was repeated many times.  For the CIC and NPF graphs, the
results are based on 6 slices, each randomized 100 times.  For the Minkowski
functional graphs, only 10 randomizations for each of the 6 slices was used.
The error bars shown on the graphs displaying our results for the CIC and NPF
statistics represent the standard deviation resulting from this procedure and
not the standard deviation of the mean.  The reason for this choice is that,
since only one data slice will be available, the error in the observations will
be approximated by the standard deviation of a single simulation rather than by
the standard deviation of the mean.

\begin{figure}
\epsfxsize=3.3in \epsfbox{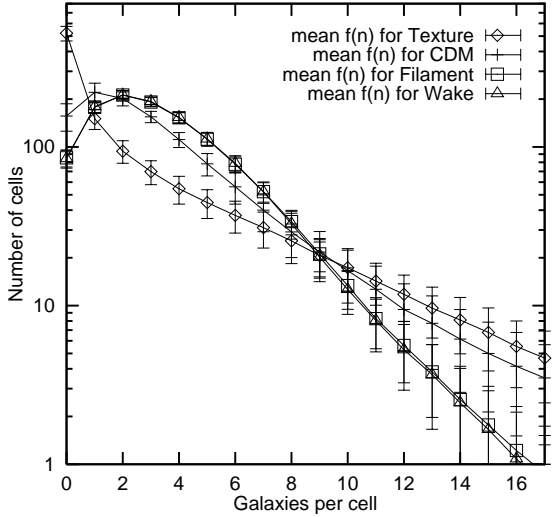}
\caption{Same as Figure 2, but for a survey with a greater limiting magnitude 
of $R = 23$.}
\end{figure}

\begin{figure}
\epsfxsize=3.3in \epsfbox{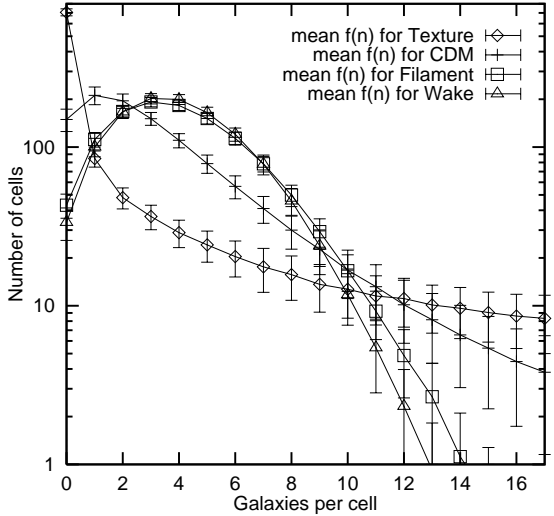}
\caption{Same as Figure 2, but for a survey with the specifications of the LZT 
survey ($\Delta z = 0.01$ and $R = 23$).}
\end{figure}

{}Figures 2 - 7 depict the CIC statistic for the four models studied in this
paper.  Figure 2 shows the results for the limiting magnitude and redshift
errors which are realized in the prototype UBC-NASA Multi-Narrowband Survey:  $R
= 20.4$ and $\Delta z = 0.03$.  Focusing on the region corresponding to a low
number of galaxies per cell, it is clear that already this prototype experiment
can yield first results of interest to cosmology.  A cosmological model like the
texture toy model can be clearly differentiated from string models and from a
standard CDM theory.  The difference between the string and CDM models, however,
is not significant.  The basic reason why already the prototype survey will be
able to distinguish between the most extreme version of large-scale structure
formation models is because the redshift error corresponds to a distance smaller
than the separation of the dominant structures in our models, which is the
comoving Hubble radius at $t_{eq}$.

As can be seen by comparing Figures 2 and 3, improving the redshift accuracy for
a fixed magnitude limit (which should be realized when the survey is complete)
greatly improves the ability of the CIC statistic to discriminate between the
different models.  In this case, the texture, string and CDM models are all
three clearly distinguishable.  As an interesting side result, we note that the
string filament and string wake models give almost identical results in spite of
the radically different structure morphology they predict.  Since it is unclear
which of these string toy models corresponds more closely to an actual cosmic
string model with HDM, the fact that they produce very similar CIC statistics is
encouraging since it means that the uncertainty in the string evolution factors
out.

\begin{figure}
\epsfxsize=3.3in \epsfbox{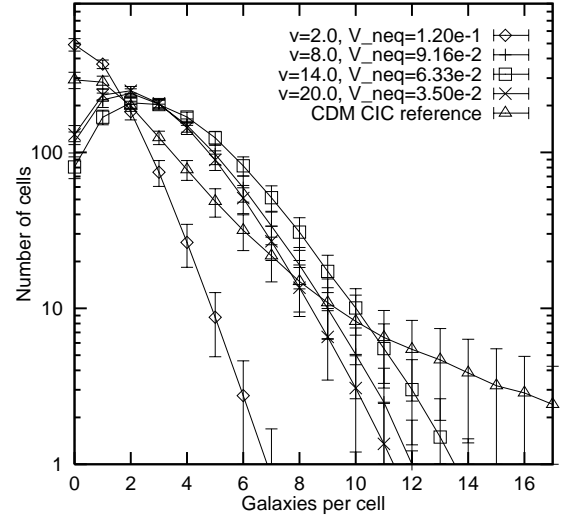}
\caption{CIC statistics for different numbers $v = \nu_2$ of wakes per Hubble 
volume ($\nu$). For comparison, the CDM curve is also shown. The variable 
$V_{neq}$ is proportional to $G \mu$.}
\end{figure}

\begin{figure}
\epsfxsize=3.3in \epsfbox{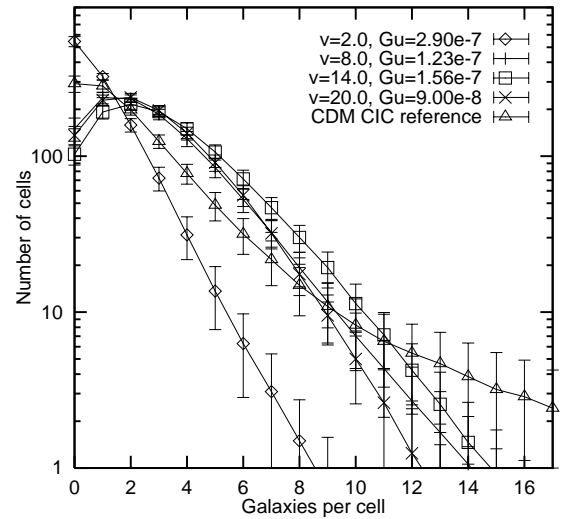}
\caption{CIC statistics for different numbers of filaments per Hubble volume 
($\nu$). For comparison, the CDM curve is also shown.}
\end{figure}

Obviously, the power of the CIC to differentiate between the models also gets 
larger by increasing the limiting magnitude (which can be realized by using a 
larger telescope or by improving the sensitivity of the CCD cameras used). 
Figure 4 shows the results which can be obtained for limiting magnitude 
$R = 23$ and redshift error $\Delta z = 0.03$, Figure 5 depicts the results 
for the target values of the LZT project, $R = 23$ and $\Delta z = 0.01$.

{}Figures 6 and 7, respectively, show how the string wake and string filament
curves vary when changing the parameters of the string models (the string
filament and wake models of Figs.  2 - 5 were run for $\nu_2 = 10$).  As the
number $\nu_2$ of structures per volume $t^3$ increases, the mass per unit
length $\mu$ must be reduced if the total clustered mass is to remain the same.
For comparison, the results for the standard CDM models are also depicted.  It
is obvious that changing the number $\nu_2$ over one order of magnitude leads to
a substantial change in the CIC curve.  For a fixed number $n$ of galaxies, the
change in the amplitude of the CIC curves between the string results for $\nu_2
= 2$ and $\nu_2 = 20$ is comparable to the difference between the CDM and string
curves.  Nevertheless, we see by inspection that the shape of the string curves
does not change substantially when $\nu_2$ varies.  Thus, by focusing on the
shape of the CIC curves over the entire range of $n$ rather than on the
amplitude for fixed $n$, one can differentiate between the theories independent
of the value of $\nu_2$.  One way to quantify the information about the shape of
the curve is to focus on its slope in the region between the value $n = n_{peak}
+ 2$ (where $n_{peak}$ is the value of $n$ corresponding to the peak of the CIC
curve; the value $n_{peak} + 2$ was chosen in order to eliminate bound ary
effects at $n = 0$) and $n = n_{1/4}$ (where $n_{1/4}$ is the value of $n$ at
$1/4$ the maximal height on the logarithmic plot).  For simulations with $R =
23$ and $\Delta z = 0.01$, the slope varies between 0.94 and 0.51 in the string
models, is about 0.31 in the CDM model and 0.11 in the texture theory.

\begin{figure}
\epsfxsize=3.3in \epsfbox{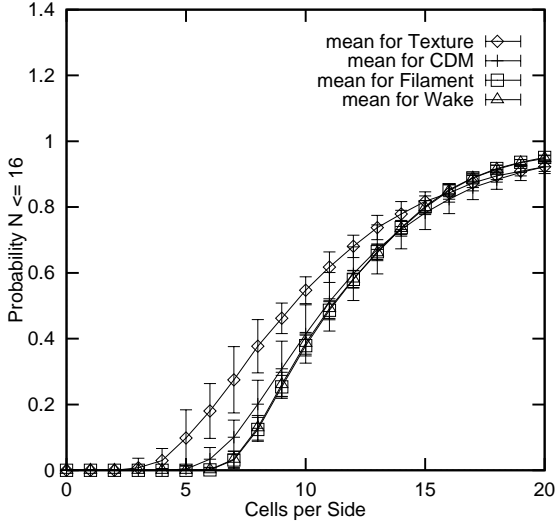}
\caption{Cumulative NPF statistic (for $N < 17$) evaluated for the 
specifications of the UBC-NASA survey for the four models considered here. 
The horizontal axis is inversely proportional to the length of an individual 
sample volume.}
\end{figure}

\begin{figure}
\epsfxsize=3.3in \epsfbox{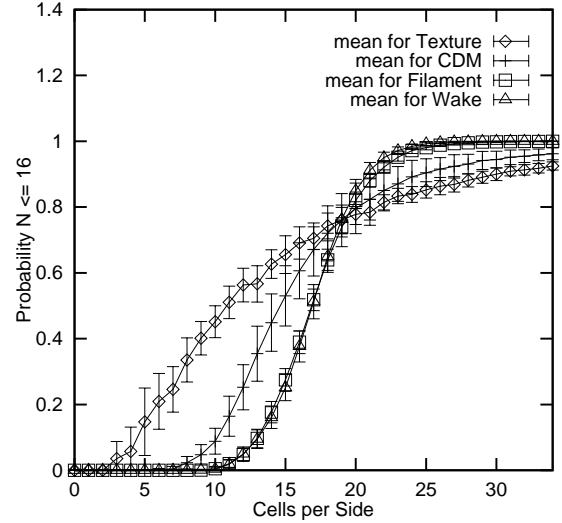}
\caption{Cumulative NPF statistic (for $N < 17$) evaluated for the 
specifications of the LZT survey.}
\end{figure}

As an interesting and unexpected side result of our study we mention the fact
that for fixed value of $\nu_2$, the string filament and string wake curves are
essentially indistinguishable.  This implies that some of the uncertainties in
the cosmic string theory of structure formation (see Section 2) cancel out when
considering the CIC statistic, at least to the accuracy studied here.

The differences in the CDM, string and texture model CIC curves stem from the
different morphologies which leads to different strengths of clustering on small
scales.  Since we model textures as spherically symmetric overdensities with a
Gaussian density profile, there is a high degree of central clustering which
leads to a large tail in the CIC curve.  On the other hand, there are also large
regions empty of galaxies which leads to the large peak at $n = 0$, where $n$ is
the number of galaxies per cell.  In contrast, the string model leads to
filamentary and planar clustering, and thus the predicted tail in the CIC curve
is small.  It is well known that nonlinear evolution (Gott et al.  1987, Park \&
Gott 1991) in the CDM model leads to a meat ball topology and hence to a
significant tail in the CIC curve.

\begin{figure}
\epsfxsize=3.3in \epsfbox{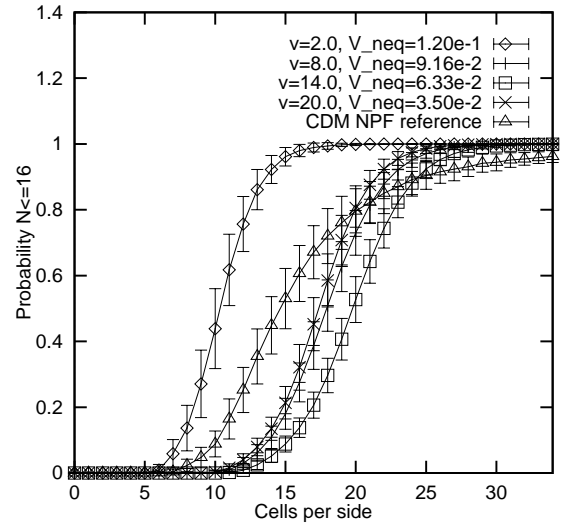}
\caption{Cumulative NPF statistic (given the specifications of the previous 
figure) in the string wake model for different values of the number of wakes 
per Hubble volume.}
\end{figure}

{}Figures 8 and 9 show the cumulative NPF statistic as a function of radius for
the specifications of the UBC-NASA and LZT surveys, respectively.  The four
curves represent the same four models for which the CIC statistics were shown in
previous figures.  The distinguishing power of the statistic was largest for the
value $N = 16$, the value used in Figures 8 and 9.  The conclusions are very
similar to those obtained by means of the CIC statistic:  with the
specifications of the current UBC-NASA survey, the texture model can already be
differentiated in a statistically significant way from the other models.  Using
conservative specifications of the LZT telescope currently under construction,
the string, texture and CDM models all give very different curves.  The string
wake and string filament models remain indistinguishable.  Figure 10 shows how
the NPF changes as the number $\nu_2$ of wakes per volume $t^3$ is varied (the
simulations used for Fig.  10 are for the specifications of the LZT survey).

\begin{figure}
\epsfxsize=3.3in \epsfbox{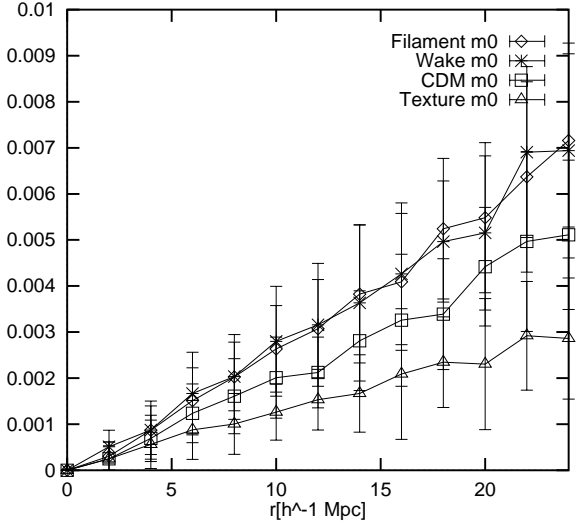}
\caption{The Minkowski functional $M_0$ for a LMT survey with specifications
$R = 20.4$ and $\Delta z = 0.01$. Note that the larger the value of $r$, the 
smaller the effective density threshold.}
\end{figure}

\begin{figure}
\epsfxsize=3.3in \epsfbox{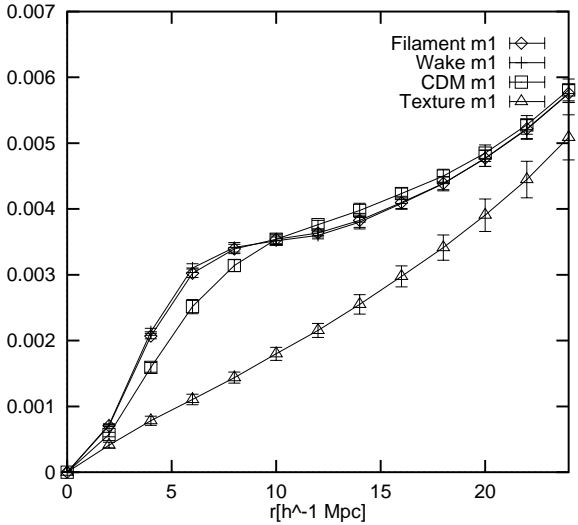}
\caption{Same for $M_1$.}
\end{figure}

To obtain the best differentiation between the various models (independent of
$\nu_2$) it is again advisable to look at the slope of the curve rather than at
the amplitude for a fixed cell volume.  The value of $\Delta x$ (where $x$ is
the number of cells per side), which is a measure of the inverse of the slope,
taken in the range over which the cumulative NPF increases from 0.2 to 0.8,
varies between 4 and 5.7 for the string models, whereas it is approximately 8.7
for the CDM simulation and 15.5 for the texture model.

Figures 11 - 14 show the four Minkowski functionals $M_0 - M_3$ evaluated for
all four models in the case of the specifications of the UBC-NASA survey with
improved redshift errors ($R = 20.4$ and $\Delta z = 0.01$).  The power of the
functionals $M_1 - M_3$ (in particular $M_1$ and $M_2$) to differentiate between
the models is comparable if not greater than the power of either the CIC or NPF
statistics.  In fact, using $M_2$ the UBC-NASA survey already offers the
potential of separating the predictions of string, texture and CDM models.

\begin{figure}
\epsfxsize=3.3in \epsfbox{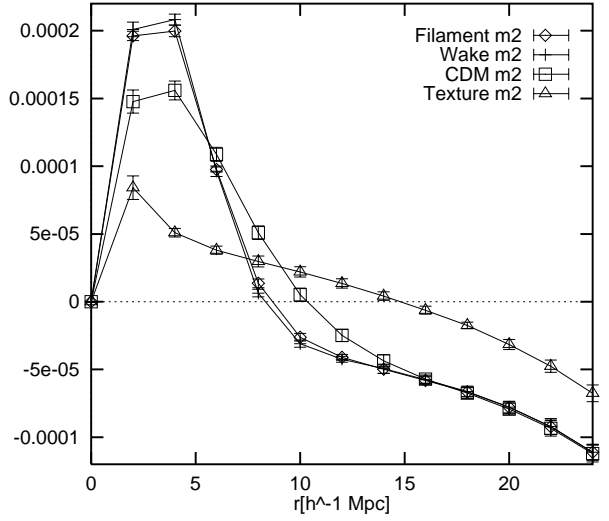}
\caption{Same for $M_2$.}
\end{figure}

\begin{figure}
\epsfxsize=3.3in \epsfbox{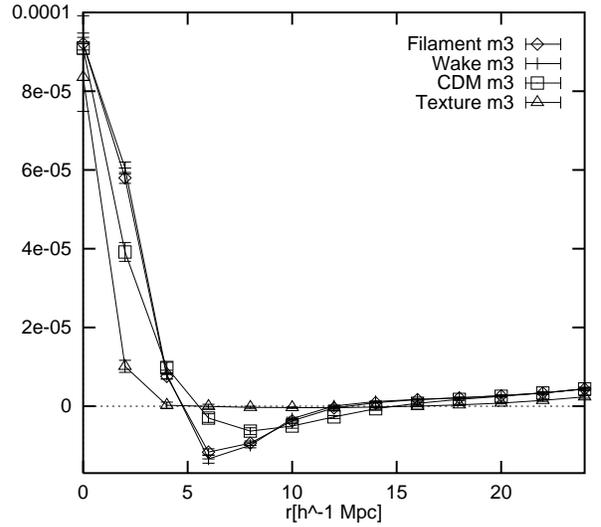}
\caption{Same for $M_3$.}
\end{figure}   

\section{Discussion and Conclusions} 

In this paper we have investigated the potential of the UBC-NASA and the LZT
liquid mirror telescope redshift surveys to be able to differentiate between the
predictions of various models of structure formation, in spite of the relatively
large redshift uncertainties inherent to the survey strategy.  We focused on the
use of topological statistics to distinguish between models with identical power
spectra but different phase structure, resulting in different morphology of the
predicted large-scale structure in the Universe.

One of our main results is that the current UBC-NASA liquid mirror telescope
survey will already be able to yield first interesting results.  In particular,
using counts in cell (CIC), N-galaxy probability functions and Minkowski
functionals, the predictions of the texture model can be separated from those of
other models with a high statistical significance.  We have seen that an
improvement in the redshift accuracy by a factor of 3 can yield a substantial
increase in the ability of the survey to distinguish between different models.
With the current limiting magnitude of $R = 20.4$ but with an improved redshift
error of $\Delta z = 0.01$, the UBC-NASA survey will be able to yield results by
means of which (making use of CIC, NPF and Minkowski functional statistics) the
cosmic string models, texture models and standard CDM models can all be
differentiated.  In fact, the resolution power is larger than what can be
achieved with $R = 23$ and $\Delta z = 0.03$.  We have seen that with a survey
having the target par ameters of the LZT project ($R = 23$ and $\Delta z = 0.01$
a lot can be learned about the origin of structure in the Universe.

A very promising statistic appears to be $M_2$, the third Minkowski functionals
which measures the integrated mean curvature.  With $M_2$, the current UBC-NASA
survey already has the potential of discriminating between string, texture and
CDM models.  It is interesting that $M_2$ appears to be a more powerful
statistic than the widely used genus statistic $M_3$.

There are, however, some important caveats concerning this study.  The
topological defect models we have analyzed are {\it toy models} and not actual
simulations which include the complete nonlinear defect and gravitational
dynamics (such simulations are at the moment out of the range of the
capabilities).  Our toy models are intended to capture the crucial morphological
differences in the predicted large-scale structure between the models,
differences which will be imprinted on the CIC, NPF and Minkowski functional
statistics which are sensitive to the morphology of structure.  Our defect toy
models to not include any super-Hubble scale correlations.  Neither do they
include correlations of the defect distribution at different Hubble times (such
correlations would be expected to be quite important in string models and
probably less important in the texture model).  We have neglected the effects of
small-scale fluctuations (imprinted before $t_{eq}$) which can be very important
(in particular in a cosmic string mo del with cold dark matter (Albrecht \&
Stebbins 1992)), and we have replaced the nonlinear dynamics of the structures
seeded by an individual defect by an ad-hoc distribution of galaxies within
these structures motivated by calculations using the Zel'dovich approximation,
thus omitting extra nonlinear gravitational clustering.  It would be interesting
to redo the analyses of this paper using improved simulations of defect models.
It would also be interesting to study the potential of the LZT project to
differentiate between the predictions of different CDM models (open CDM,
${\Lambda}CDM$, mixed dark matter models).  Work on these issues is in progress.

\begin{center}{\Large  Acknowledgments}\end{center}

We are grateful to Martin G\"otz for running the CDM simulation for us.  We
thank Ed Bertschinger for the use of his code (Bertschinger \& Gelb 1991),
Thomas Buchert and collaborators for making their Minkowski functional codes
publically available, and Gerry Guralnik for granting us access to the computing
facilities of the Brown Center for Scientific Computation.  One of us (R.B.)
wishes to thank Bill Unruh for hospitality at UBC in Vancouver over the period
of several visits when some of the work reported here was carried out, and
Martin Kerscher for comments on the draft.  This work was supported in part by
the U.S.  Department of Energy under Contract DE-FG0291ER40688.  D.M.  was
supported in part by a Brown University Royce Fellowship and by the NASA Space
Grant Program. P.H. is supported by grants from the Natural Sciences and
Engineering Research Council of Canada.


\begin{thebibliography}{}

\bibitem{Aguirre2} Aguirre, A., 1995, Senior thesis, Brown Univ. (unpublished).
\bibitem{Aguirre} Aguirre, A., \&  Brandenberger, R., 1995, Int. J. Mod. Phys.,
	D4, 711; astro-ph/9505031.
\bibitem{AlbSteb} Albrecht, A. \& Stebbins, A., 1992, Phys. Rev. Lett., 68, 
	2121.
\bibitem{AT} Albrecht, A. \&  Turok, N., 1989,  Phys. Rev., D40, 973.
\bibitem{AS} Allen, B. \& Shellard, E.P.S., 1990, Phys. Rev. Lett., 64, 119.
\bibitem{BBKS} Bardeen, J., Bond, J., Kaiser, N., \& Szalay, A., 1986, Ap. J., 
	304, 15.
\bibitem{} Bardeen, J., Steinhardt, P. \& Turner, M., 1983, Phys. Rev., D28, 
	1809.
\bibitem{BB} Bennett, D. \& Bouchet, F., 1988, Phys. Rev. Lett., 60, 257.
\bibitem{BG} Bertschinger, E. \& Gelb, J., 1991, Computers in Physics, 5, 164.
\bibitem{Borra} Borra. E.F., Content, R., Girard, L., Szapiel, S., 
	Tremblay, L.-M., \& Boily, E., 1992, Ap. J., 393, 829.
\bibitem{RMP85} Brandenberger, R., 1985, Rev. Mod. Phys., 57, 1.
\bibitem{} Brandenberger, R., 1991, Phys. Scripta, T36, 114.
\bibitem{RBrev} Brandenberger, R., 1994, Int. J. Mod. Phys., A9, 2117.
\bibitem{} Brandenberger, R., Kaiser, N., Schramm, D. \&  Turok, N., 1987a, 
	Phys. Rev. Lett., 59, 2371.
\bibitem{} Brandenberger, R., Kaiser, N. \&  Turok, N., 1987b, Phys. Rev., 
	D36, 2242.
\bibitem{} Chibisov, G. \& Mukhanov, V., 1980, `Galaxy Formation and
	Phonons,' Lebedev Physical Institute Preprint No. 162.
\bibitem{} Chibisov, G. \& Mukhanov, V., 1982, MNRAS 200, 535.
\bibitem{Davis} Davis, R., 1987, Phys. Rev., D35, 3705.
\bibitem{CFA2} de Lapparent, V., Geller, M. \&  Huchra, J., 1986, Ap. J.
	(Lett.), 302, L1.
\bibitem{Gooding} Gooding, A., Spergel, D. \& Turok, N., 1991, Ap. J. Lett.,
	372, L5.
\bibitem{nonlinear} Gott III, J.R., Weinberg, D. \& Melott, A., 1987, Ap. J.,  319, 1.
\bibitem{} Guth, A. \&  Pi, S.-Y., 1982,  Phys. Rev. Lett.,  49, 110.
\bibitem{Hamilton} Hamilton, A., Saslaw, W. \&  Thuan, T., 1985, Ap. J., 297, 
	37.
\bibitem{} Hawking, S., 1982, Phys. Lett. 115B, 295.
\bibitem{Hickson2} Hickson, P., Borra, E. F., Cabanac, R., Content, R., 
	Gibson, B. K., \& Walker, G. A. H., 1994, Ap. J., 436, L201.
\bibitem{LZT} Hickson, P., Borra, E. F., Cabanac, R., Chapman, S., 
	de Lapparent, V., Mulrooney, M. \& Walker, G. A. H., 1998,
in Advanced Technology Optical/IR Telescopes VI,  Proc. SPIE, 3352, in press.
\bibitem{Hickson3} Hickson, P., Gibson, B. K. \& Callaghan K., 1994, MNRAS, 
	267, 911.
\bibitem{Hickson1} Hickson, P. \& Mulrooney, M. K., 1998, Ap. J. (Suppl.), 
	115, 35.
\bibitem{HKrev} Hindmarsh, M. \& Kibble, T.W.B., 1995, Rept. Prog. Phys., 58, 
	477.
\bibitem{Kerscher} Kerscher, M. et al., 1997, MNRAS, 284, 73; astro-ph/9606133.
\bibitem{TK76} Kibble, T.W.B., 1976, J. Phys., A9, 1387.
\bibitem{Koo} Koo, D., 1985, Astron. J., 90, 418.
\bibitem{Lanzetta} Lanzetta, K., Yahil, A. \& Fernandez-Soto, A., 1996, Nature, 
	381, 759.
\bibitem{} Leese, R. \&  Prokopec, T., 1991, Phys. Rev., D44, 3749.
\bibitem{} Lukash, V., 1980,  Pis'ma Zh. Eksp. Teor. Fiz., 31, 631.
\bibitem{Maddux} Maddox, S., 1998, in Proceeding of Potsdam Cosmology Workshop 
	"Large Scale Structure: Tracks and Traces", in press.; 
	astro-ph/9711015.
\bibitem{Mecke1} Mecke, K., Buchert, T. \& Wagner, H., 1994, Astron. 
	Astrophys. 288, 697.
\bibitem{Mecke2} Mecke, K. \& Wagner, H., 1991, J. Stat. Phys., 64, 843.
\bibitem{} Mukhanov, V. \& Chibisov, G., 1981, JETP Lett., 33, 532.
\bibitem{} Park, C. \&  Gott III, J.R., 1991, Ap. J., 378, 457.
\bibitem{PST} Pen, U.-L., Seljak, U. \&  Turok, N., 1997, Phys. Rev. Lett., 
	79, 1611.
\bibitem{HDM3} Perivolarapoulos, L., Brandenberger, R. \& Stebbins, A., 1990, 
	Phys. Rev., D41, 1764.
\bibitem{Press} Press, W., 1980, Phys. Scr., 21, 702.
\bibitem{Prokopec} Prokopec, T., 1991, Phys. Lett., 262B, 215.
\bibitem{Saslaw} Saslaw, W., 1989, Ap. J., 341, 588.
\bibitem{Buchert} Schmalzing, J. \& Buchert, T., 1997,  Ap. J. (Lett.), 482, 
	L1;  astro-ph/9702130.
\bibitem{COBE} Smoot, G. et al., 1992, Ap.J. (Lett.), 396, L1.  
\bibitem{wake} Silk, J. \&  Vilenkin, A., 1984, Phys. Rev. Lett., 53, 1700.
\bibitem{} Starobinskii, A.,1982, Phys. Lett., 117B, 175.
\bibitem{Turok} Turok, N., 1989, Phys. Rev. Lett., 63, 2625.
\bibitem{deficit} Vilenkin, A., 1981, Phys. Rev., D23, 852.
\bibitem{Vil85} Vilenkin, A., 1985, Phys. Rep., 121, 263.
\bibitem{HDM1} Vilenkin, A. \& Shafi, Q., 1983, Phys. Rev. Lett., 51, 1716.
\bibitem{VSbook} Vilenkin, A. \& Shellard, E.P.S., 1994, Strings and Other 
	Topological Defects. Cambridge Univ. Press, Cambridge.
\bibitem{ZLB} Zanchin, V., Lima, J.A.S., \& Brandenberger, R., 1996, Phys. 
	Rev.,  D54, 7219; astro-ph/9607062.
\bibitem{Zeld} Zel'dovich, Ya. B., 1970, Astron. Astrophys., 5, 84 (1970).
\bibitem{} Zucca, E.  et al., 1997, Astron. Ap., 326, 477.

 
\end{thebibliography}
\end{document}